\documentclass[aps,prc,preprint,groupedaddress,showpacs]{revtex4}
\usepackage{epsfig}
\usepackage{amsmath}
\usepackage[hyperref]{hyperref}

\begin{document}

\title{A further analysis of $\Theta^+$ production in $\gamma+D \to \Lambda+n+K^{+}$ reaction}

\author{V. Guzey}

\email{vadim.guzey@tp2.rub.de}
\affiliation{Institut f{\"u}r Theoretische Physik II, Ruhr-Universit{\"a}t
Bochum, D-44780 Bochum, Germany}

\begin{abstract}

We analyse $\Theta^+$ production in the $\gamma+D \to \Lambda +n +K^+$ reaction and
study the dependence of the $\gamma+D \to \Lambda +n +K^+$ differential
cross section on the $n K^+$ invariant mass and on the momentum of the
final neutron $p_n$. We examine the important role 
of the interference between the 
signal and background contributions to the $\gamma+D \to \Lambda +n +K^+$
amplitude in the extraction of the $\Theta^+$ signal from the $\gamma+D \to \Lambda +n +K^+$ cross section.
We demonstrate that as a result of the
cancellation between the interference and signal contributions, the
$\Theta^+$ signal almost completely washes out after the integration over $p_n$.
This is consistent with the CLAS conclusion that no statistically 
significant structures in the analysis of the $\gamma+D \to \Lambda +n +K^+$ 
reaction were observed.
Therefore, there is no disagreement between the theory and the experiment
and
the CLAS result does not refute the existence of the $\Theta^+$.

\end{abstract}

\pacs{13.60.Rj, 14.20.Jn, 25.20.Lj}
\preprint{RUB-TP2-06/2006}

\maketitle

\section{Introduction}
\label{sec:intro}

Three and a half years have passed since the announcement of the discovery
of an exotic pentaquark baryon, the $\Theta^+$, by the LEPS collaboration~\cite{Nakano:2003qx}.
This announcement has renewed the interest and enthusiasm in hadron spectroscopy: 
The original 
LEPS publication was followed by tens of measurements
aiming to confirm or refute the LEPS result and by hundreds of
theoretical analyses of properties of the $\Theta^+$, possible
mechanisms of its production and many other related issues,
see~\cite{Hicks:2005gp} for a review of
the experimental situation in 2005. 
Over the time, the experimental status of the $\Theta^+$ has changed 
dramatically:
Many initial experiments confirming the $\Theta^+$ were followed by 
mostly negative results on the $\Theta^+$ search.
At the moment, the only experiment, which officially confirmed its own
 initial claim of the 
$\Theta^+$ discovery, is the measurement of the 
$K^+ Xe \to K^0 p Xe^{\prime}$ reaction
by DIANA collaboration~\cite{Barmin:2006we}.
In summary, if the $\Theta^+$ exists, its exotic properties make it very difficult
to unambiguously establish its existence. All indirect measurements used so far proved 
to be inefficient and marred by significant experimental and theoretical uncertainties.
Most likely, only future dedicated experiments using kaon beams 
(note the still standing positive result of DIANA~\cite{Barmin:2006we} using the
kaon beam) will be able
to answer the question whether the $\Theta^+$ exists or not.

The properties of the $\Theta^+$ and other members of the antidecuplet, 
the flavor SU(3)
multiplet containing the $\Theta^+$, were first predicted in the chiral
quark soliton model by Diakonov, Petrov and Polyakov in 1997~\cite{Diakonov:1997mm}.
The predicted positive strangeness of the $\Theta^+$ indicates that in the language
of the quark model, the $\Theta^+$ has the minimal structure $u u d d {\bar s}$, 
i.e.~it is an exotic pentaquark baryon. Another remarkable property of the $\Theta^+$,
which makes it ''doubly exotic'', is the predicted small total width of the $\Theta^+$,
$\Gamma_{\Theta} < 15$ MeV~\cite{Diakonov:1997mm}. In reality, the total width is possibly
even smaller: The recent theoretical estimates range between 2-5 MeV~\cite{Diakonov:2005ib,Lorce:2006nq} and 1 
MeV~\cite{Arndt:2003xz,Cahn:2003wq,Sibirtsev:2004bg}. The experimental upper bound
on $\Gamma_{\Theta}$ is usually given by the spectrometer resolution,
which is of the order of 5-10 MeV. However, in some cases, more stringent 
constraints can be derived. For instance, the recent DIANA analysis reports
$\Gamma_{\Theta}=0.36 \pm 0.11$ MeV~\cite{Barmin:2006we}.

The small value of $\Gamma_{\Theta}$, the unknown mechanism of production of the 
$\Theta^+$ (the necessity to impose various cuts on the data in order to enhance the
signal, which dramatically
reduces the statistics) and generally rather small
cross sections of the $\Theta$ production (at the nanobarn level) make all analyses attempting to
extract the $\Theta^+$ signal model-dependent and inconclusive.
Naturally, this does not necessarily mean that the $\Theta^+$ does not exist. 
The majority of the experiments simply might not have enough sensitivity (enough statistics)
 to observe the elusive $\Theta^+$.

In this work, we consider protoproduction of
the $\Theta^+$ on deuterium in the reaction $\gamma+D \to \Lambda +n +K^+$.
While we comprehensively studied this reaction earlier~\cite{Guzey:2004jq},
the very recent CLAS measurement~\cite{Niccolai:2006td} 
compels us to perform further studies. In particular, in order to make a better 
comparison  between our theoretical predictions and the experimental results,
we analyse in detail the differential cross section of the $\gamma+D \to \Lambda +n +K^+$
process as function of the $n K^+$ invariant mass and of the momentum of the
final neutron, $p_n$. We discuss the important role of the interference between the 
signal and background and examine its $p_n$-dependence.
We demonstrate that after integrating over $p_n$ and the photon
energy as was done in the CLAS analysis~\cite{Niccolai:2006td}, one does not 
expect any significant structures associated with the $\Theta^+$ in the 
$\gamma+D \to \Lambda +n +K^+$ cross section. 
This is consistent with the CLAS analysis, which reports no statistically 
significant structures~\cite{Niccolai:2006td}.
Therefore, there is no disagreement between the theory and the experiment.
The CLAS result does not refute the existence of the $\Theta^+$.

\section{Interference between the background and the signal and the size of
the $\Theta^+$ signal}
\label{sec:signal}

In order to make our presentation self-contained, we shall 
repeat in detail the derivation of key expressions from our original
analysis~\cite{Guzey:2004jq}.

\subsection{The signal amplitude}

\begin{figure}[t]
\begin{center}
\epsfig{file=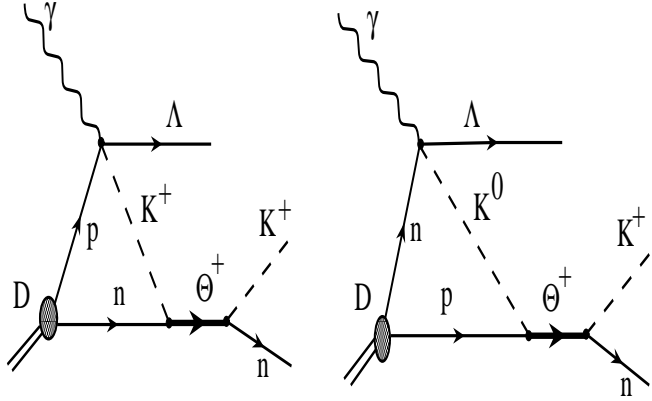,width=10cm,height=10cm}
\vskip -3cm
\caption{The assumed mechanism of $\Theta^+$ production in the
$\gamma+D \to \Lambda +n +K^+$ reaction.}
\label{fig:signal}
\end{center}
\end{figure}

We assume that the dominant mechanism of the $\Theta^+$ production in the
reaction $\gamma+D \to \Lambda +n +K^+$ is given by the Feynman graphs 
presented in Fig.~\ref{fig:signal}. The corresponding scattering amplitude 
reads
\begin{eqnarray}
{\cal A}_{{\rm S}}&=&-i \int \frac{d^4 k}{(2 \pi)^4}\bar{u}(p_n)\hat{\Gamma}_{\Theta}
\frac{\hat{p}_{\Theta}+M_{\Theta}}{p_{\Theta}^2-M_{\Theta}^2+i\Gamma_{\Theta}M_{\Theta}}\hat{\Gamma}_{\Theta} \nonumber\\
&\times&
\frac{\hat{k}+m_N}{k^2-m_N^2+i0} \frac{1}{(p_{\Theta}-k)^2-m_K^2+i0} 
\bar{u}(p_{\Lambda}) \hat{\Gamma}_{\Lambda}^{p+n}  
\frac{\hat{p}_D-\hat{k}+m_N}{(p_D-k)^2-m_N^2+i0} \hat{\Gamma}_{D} \,,
\label{eq:a_signal}
\end{eqnarray}
where $p_n$ is the momentum of the final neutron; $p_{\Lambda}$ is the momentum of
 the $\Lambda$; $p_{\Theta}$ is the momentum of the virtual $\Theta^+$;
$k$ is the momentum of the spectator nucleon in the loop;
the $\hat{\Gamma}_{\Theta}$ vertex describes the $\Theta^+ \to n K^+$ transition;
the $\hat{\Gamma}_{\Lambda}^{p+n}$ vertex describes the sum of 
the $\gamma+p \to \Lambda+K^+$ and  $\gamma+n \to \Lambda+K^0$
transitions; the $\hat{\Gamma}_{D}$ vertex describes the $D \to pn$ transition.
For brevity, we do not write
explicitly spinor polarization indices.

The loop integration in Eq.~(\ref{eq:a_signal}) receives its dominant 
contribution from the integration region, when the spectator nucleon 
and the virtual kaon tend to be on the mass-shell. 
Then, the dominating imaginary part of the loop integral
can be evaluated through the discontinuity of the loop integral
using Cutkosky cutting rules
\begin{eqnarray}
&&2\, Im \left(i \int  \frac{d^4 k}{(2 \pi)^4}\frac{1}{k^2-m_N^2+i0} \frac{1}{(p_{\Theta}-k)^2-m_K^2+i0}\right)= \nonumber\\
&&(2 \pi i)^2 \int \frac{d^4 k}{(2 \pi)^4}
\delta(k^2-m_N^2) \delta\left((p_{\Theta}-k)^2-m_K^2\right)
\Theta(k^0) \,\Theta(p_{\Theta}^0-k^0) \,.
\end{eqnarray}

As we discussed in the Introduction,
$\Gamma_{\Theta}$ is very small. Hence, 
the signal amplitude is non-vanishing only near $p_{\Theta}^2 \approx
M_{\Theta}^2$ and one can approximately write
\begin{equation}
\frac{1}{p_{\Theta}^2-M_{\Theta}^2+i\Gamma_{\Theta}M_{\Theta}} \approx
-i\, \frac{\Gamma_{\Theta}M_{\Theta}}{(p_{\Theta}^2-M_{\Theta}^2)^2+(\Gamma_{\Theta}M_{\Theta})^2} \,.
\end{equation}
Therefore, the signal scattering amplitude can be written as
\begin{eqnarray}
{\cal A}_{{\rm S}}&=&\int \frac{d^4 k}{(2 \pi)^4}\bar{u}(p_n)\hat{\Gamma}_{\Theta}
u(p_{\Theta}) \, \bar{u}(p_{\Theta})\hat{\Gamma}_{\Theta} u(k) \,
\bar{u}(p_{\Lambda}) \hat{\Gamma}_{\Lambda}^{p+n} u(p_D-k)\,\frac{\Gamma_{\Theta}M_{\Theta}}{(p_{\Theta}^2-M_{\Theta}^2)^2+(\Gamma_{\Theta}M_{\Theta})^2}
 \nonumber\\
&\times &\bar{u}(p_D-k) \bar{u}(k) \frac{\hat{\Gamma}_{D}}{(p_D-k)^2-m_N^2+i0}
\delta(k^2-m_N^2) \delta\left((p_{\Theta}-k)^2-m_K^2\right)
\nonumber\\
&\times &\Theta(k^0) \,\Theta(p_{\Theta}^0-k^0) \,,
\label{eq:a_signal2}
\end{eqnarray}
where $p_D-k$ is the momentum of the interacting nucleon in the loop.
It is important to note that the resulting signal scattering amplitude 
has an overall plus sign. As will be shown below, 
this will lead to the destructive  
interference between the signal and the background amplitudes. 
In our original work~\cite{Guzey:2004jq},
it was erroneously derived that the sign should be negative. However, this
mistake was not important there since the role of the interference 
between the two amplitudes was very small in the kinematics
considered in~\cite{Guzey:2004jq}.

The factor $\Gamma_{\Theta}M_{\Theta}/[(p_{\Theta}^2-M_{\Theta}^2)^2+(\Gamma_{\Theta}M_{\Theta})^2]$ in Eq.~(\ref{eq:a_signal2}) forces the virtual $\Theta^+$ on its mass-shell.
This means that one can neglect the off-shellness
of the $\Theta^+$ in the $\Theta^+ \to N K$ vertex. Therefore,
the corresponding factor in Eq.~(\ref{eq:a_signal2}) depends only on the 
involved masses and $\Gamma_{\Theta}$,
\begin{equation}
\bar{u}(p_n)\hat{\Gamma}_{\Theta}
u(p_{\Theta}) \, \bar{u}(p_{\Theta})\hat{\Gamma}_{\Theta} u(k)=
\frac{8 \pi M_{\Theta}^3 \Gamma_{\Theta}}{\lambda^{1/2}(M_{\Theta}^2,m_N^2,m_K^2)} 
\,,
\label{eq:onshell}
\end{equation}
where $\lambda(x,y,z)$ is the so-called
triangular kinematic function, $\lambda(x,y,z)=(x-y-z)^2-4\, y^2\,z^2$.
In the derivation of Eq.~(\ref{eq:onshell}) we used the textbook relation
between the $1 \to 2$ decay amplitude and the corresponding decay width and
the fact that $\Gamma_{\Theta^+ \to nK^+}=(1/2) \Gamma_{\Theta}$.

In the vertex describing the $\gamma+N \to \Lambda +K$ transition in
Eq.~(\ref{eq:a_signal2}), the interacting nucleon is off-shell. However,
since the effect of the off-shellness is small, we shall ignore it and
use for the $\gamma+N \to \Lambda +K$  scattering amplitudes their on-shell
expressions~\cite{MAID}. Moreover, since the sensitivity of our results
to the fine details of the $\gamma+N \to \Lambda +K$ transition is
not large and since the nuclear wave function suppresses the contribution
of large nucleon momenta,
we assume that the scattering takes place on the nucleon at rest.
At this point, it is convenient to introduce the following
short-hand notation
\begin{equation}
\bar{u}(p_{\Lambda}) \hat{\Gamma}_{\Lambda}^{p+n} u(p_D-k)=V_{\Lambda}^{p+n}(E_{\gamma},t) \,,
\label{eq:short_hand}
\end{equation}
which demonstrates that the $\gamma+N \to \Lambda +K$ amplitude depends only on
the photon energy $E_{\gamma}$ and 
$t=(p_{\gamma}-p_{\Lambda})^2$.

 One has to point out that the $\gamma+p \to \Lambda +K^+$
and $\gamma+n \to \Lambda +K^0$
scattering amplitudes, which correspond to the left and right
hand side figures in Fig.~\ref{fig:signal}, respectively,
 enter Eq.~(\ref{eq:a_signal2}) with the relative plus sign. This
is a consequence of the observation that only the isospin-zero part of the
electromagnetic current contributes (we used that the $\Theta^+$ has 
isospin zero).

Further, the $D \to NN$ vertex $\hat{\Gamma}_{D}$ in Eq.~(\ref{eq:a_signal2})
can be related to the deuteron non-relativistic wave function $\psi_D$
\begin{equation}
\bar{u}(p_D-k) \bar{u}(k) \frac{\hat{\Gamma}_D}{(p_D-k)^2-m_N^2+i0}=
\sqrt{(2 \pi)^3 2 m_N}\, \psi_D(k) \,.
\end{equation}
This non-relativistic reduction means that we have to work in the laboratory 
(target rest) frame. In our work, we used the Paris deuteron wave 
function~\cite{Lacombe:1981eg}.

The two delta-functions in Eq.~(\ref{eq:a_signal2})
enable one to take integrals over $k^0$ and the angle between the vectors 
${\vec k}$ and ${\vec p}_{\Theta}$ (nothing depends on the azimuthal angle, so
that the integration over it simply gives $2 \pi$).

The resulting expression for the signal scattering amplitude has the following
compact form
\begin{eqnarray}
{\cal A}_{{\rm S}}&=&\frac{\Gamma_{\Theta}M_{\Theta}}{(M_{NK}^2-M_{\Theta}^2)^2+(\Gamma_{\Theta}M_{\Theta})^2}\, V_{\Lambda}^{p+n}(E_{\gamma},t)\,\frac{8 \pi M_{\Theta}^3 \Gamma_{\Theta}}{\lambda^{1/2}(M_{\Theta}^2,m_N^2,m_K^2)} \nonumber\\
&\times& \frac{\sqrt{(2 \pi)^3 2 m_N}}{16 \pi}
\int dk \frac{k}{E_k} \Theta(E_{\Theta}-E_k)\frac{\theta(E_{\gamma},t,M^2_{NK};k)}{|\vec{p}_{\Theta}|}\, \psi_D(k) \,,
\label{eq:a_signal3}
\end{eqnarray}
where  $M_{NK}$ is the invariant mass of the final $nK^+$ system,
$M_{NK}^2 \equiv p_{\Theta}^2$;
$\theta(E_{\gamma},t,M^2_{NK};k)$ denotes the $\theta$-function remaining after the angular
integration,
\begin{equation}
\theta(E_{\gamma},t,M^2_{NK};k) \equiv \Theta \left(-1 \leq  \frac{2 E_{\Theta} E_k+m_K^2-m_N^2-M^2_{NK}}{2 |\vec{p}_{\Theta}|k} \leq 1 \right) \,.
\label{eq:theta_k}
\end{equation}
In Eqs.~(\ref{eq:a_signal3}) and (\ref{eq:theta_k}), $E_k=\sqrt{k^2+m_N^2}$; $p_{\Theta}=\sqrt{(E_{\gamma}-E_{\Lambda})^2-t}$;
$E_{\Theta}=\sqrt{M_{NK}^2+(\vec{p}_{\Theta})^2}$;
$E_{\Lambda}=(t+m_D^2+2 m_DE_{\gamma}-M_{NK}^2)/(2 m_D)$.

\subsection{The background amplitude}

\begin{figure}[t]
\begin{center}
\epsfig{file=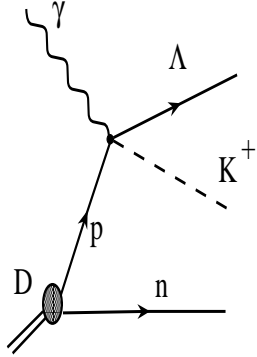,width=11cm,height=11cm}
\vskip -4cm
\caption{The main background contribution to the $\gamma+D \to \Lambda +n +K^+$ reaction in the kinematics considered in this work.}
\label{fig:bg}
\end{center}
\end{figure}

We suppose that the main background contribution to the considered
process is the $\gamma+p \to \Lambda +K^+$ scattering on the quasi-free proton,
which is depicted in Fig.~\ref{fig:bg}. Note that other backround reactions 
are also possible~\cite{Laget:2006ym}, but 
their contributions are suppressed in the kinematics considered in this work.

The background scattering amplitude corresponding to the Feynman graph in
Fig.~\ref{fig:bg} reads
\begin{eqnarray}
{\cal A}_{{\rm BG}}&=&-\bar{u}(p_{\Lambda}) \hat{\Gamma}_{\Lambda}^p
\frac{\hat{p}_D-\hat{k}+m_N}{(p_D-k)^2-m_N^2+i0} \bar{u}(p_n)\hat{\Gamma}_{D}
\nonumber\\
&=&-\sqrt{(2 \pi)^3 2 m_N} \psi_D(p_n) V_{\Lambda}^p(E_{\gamma},t) \,,
\label{eq:a_bg}
\end{eqnarray}
where $V_{\Lambda}^p(E_{\gamma},t)$ denotes the 
$\gamma+p \to \Lambda +K^+$ scattering amplitude. In the derivation of the
second line of Eq.~(\ref{eq:a_bg}), we have used the conventions and 
approximations described in the previous subsection.

The background amplitude has an overall minus sign. Therefore, its
interference with the signal amplitude is destructive. As we shall argue
below, the non-observation of the $\Theta^+$ in the CLAS measurement~\cite{Niccolai:2006td}
can be partially explained by
the non-trivial dependence of this interference on the momentum of the final
neutron $p_n$.

\subsection{The $\gamma+D \to \Lambda+n+K^{+}$ differential cross section
and the role of the interference between the background and the signal}

We would like to study the differential cross section of the 
$\gamma+D \to \Lambda+n+K^{+}$ reaction as a function of $t$, $p_n$ and $M^2_{NK}$.
For this purpose, it is convenient to present the three-body phase space of
the final particles as~\cite{Kajantie}
\begin{eqnarray}
&& R_3(\gamma\,D \to \Lambda\,n\,K^{+}) \equiv \int \frac{d^3 p_{\Lambda}}{2 E_{\Lambda}} \frac{d^3 p_{n}}{2 E_{n}}
\frac{d^3 p_{K}}{2 E_{K}} \delta^4(p_{\gamma}+p_D-p_{\Lambda}-p_{n}-p_{K})=
\nonumber\\
&& \int dM^2_{NK} \,R_2(\gamma\,D \to \Lambda\,\Theta^+) \, R_2(\Theta^+ \to n\,K^+)
\label{eq:kajantie1} \,,
\end{eqnarray}
where $R_2$ stand for the corresponding two-body  phase spaces.
A straightforward calculation gives~\cite{Kajantie}
\begin{eqnarray}
&& R_2(\gamma\,D \to \Lambda\,\Theta^+)=\frac{2\, \pi}{ 4 \,\lambda^{1/2}(s,0,m_D^2)} \,dt
\,, \nonumber\\
&&R_2(\Theta^+ \to n\,K^+)=\frac{\pi}{2}\, \frac{p_n}{E_n}\, \frac{\theta(E_{\gamma},t,M^2_{NK};p_n)}{p_{\Theta}}\, d p_n
\,,
\label{eq:kajantie2}
\end{eqnarray}
where $s=2 E_{\gamma}m_D+m_D^2$;
 the $\theta$-function $\theta(E_{\gamma},t,M^2_{NK};p_n)$ is given by Eq.~(\ref{eq:theta_k}) after
the replacement of the spectator nucleon momentum $k$ by the final
neutron momentum $p_n$.

The triple differential cross section of the $\gamma+D \to \Lambda+n+K^{+}$ reaction
has now the following form
\begin{equation}
\frac{ d \sigma}{dt d p_n d M^2_{NK}}=\frac{1}{(2 \pi)^5}\,\frac{1}{4 \,I_D} \,|{\cal A}_{{\rm BG}}+{\cal A}_{{\rm S}}|^2 \, \frac {d R_2(\gamma\,D \to \Lambda\,\Theta^+)}{dt} \, \frac{d R_2(\Theta^+ \to n\,K^+)}{d p_n} \,,
\label{eq:kajantie3}
\end{equation}
where $I_D$ is the standard flux factor evaluated for real photons impinging on the deuterium
target, $I_D=\lambda^{1/2}(s,0,m_D^2)/2$. 

As is clear from Eq.~(\ref{eq:kajantie3}), the cross section receives 
contributions from the background, interference between the background and the
signal and the signal cross sections
\begin{equation}
\frac{ d \sigma}{dt d p_n d M^2_{NK}}=\frac{ d \sigma^{{\rm BG}}}{dt d p_n d M^2_{NK}}-
\frac{ d \sigma^{{\rm I}}}{dt d p_n d M^2_{NK}}+\frac{ d \sigma^{{\rm S}}}{dt d p_n d M^2_{NK}}\,.
\label{eq:kajantie4}
\end{equation}
Evaluating $|{\cal A}_{{\rm BG}}+{\cal A}_{{\rm S}}|^2$ in Eq.~(\ref{eq:kajantie3}), 
the background, interference and signal cross sections in Eq.~(\ref{eq:kajantie4}) 
take the following form
\begin{eqnarray}
 \frac{ d \sigma^{{\rm BG}}}{dt d p_n d M^2_{NK}}& =& \frac{\pi}{4} m_N \frac{p_n}{E_n} \,|\psi_D(p_n)|^2\, \frac{ d \sigma_{\Lambda}^p}{dt} \,
\frac{\theta(E_{\gamma},t,M^2_{NK};p_n)}{p_{\Theta}} \,,
\nonumber\\
\frac{ d \sigma^{{\rm I}}}{dt d p_n d M^2_{NK}}&=&2 \,\Gamma_{\Theta}\,
\frac{\Gamma_{\Theta}M_{\Theta}}{(M^2_{NK}-M_{\Theta}^2)
+(\Gamma_{\Theta}M_{\Theta})^2} \,
\frac{M_{\Theta}^3}{\lambda^{1/2}(M_{\Theta}^2,m_N^2,m_K^2)}\, \frac{ d \sigma_{\Lambda}^I}{dt} \,S_D(t,p_n) \,,
\nonumber\\
\frac{ d \sigma^{{\rm S}}}{dt d p_n d M^2_{NK}}&=&
\Gamma_{\Theta}
\frac{\Gamma_{\Theta}M_{\Theta}}{(M^2_{NK}-M_{\Theta}^2)
+(\Gamma_{\Theta}M_{\Theta})^2} 
\frac{M_{\Theta}^5}{\lambda(M_{\Theta}^2,m_N^2,m_K^2)} \frac{ d \sigma_{\Lambda}^{p+n}}{dt} \nonumber\\
&\times&
\frac{p_n}{E_n} \frac{\theta(E_{\gamma},t,M^2_{NK};p_n)}{p_{\Theta}} S_D(t) \,.
\label{eq:main}
\end{eqnarray}
In this equation, the combinations of the elementary $\gamma +N \to \Lambda +K$ cross sections
are given by the following expressions
\begin{eqnarray}
\frac{ d \sigma_{\Lambda}^p}{dt}&=&\frac{1}{64 \,\pi\, (E_{\gamma} m_N)^2} |{\cal A}^p|^2
\,, \nonumber\\
\frac{ d \sigma_{\Lambda}^I}{dt}&=&\frac{1}{64 \,\pi \,(E_{\gamma} m_N)^2} 
\left({\cal A}^p ({\cal A}^p+{\cal A}^n)^{\ast}+({\cal A}^p)^{\ast} 
({\cal A}^p+{\cal A}^n)\right)
\,, \nonumber\\
\frac{ d \sigma_{\Lambda}^{p+n}}{dt}&=&\frac{1}{64 \,\pi\, (E_{\gamma} m_N)^2}
 |{\cal A}^p+{\cal A}^n|^2
\,, 
\label{eq:lambda}
\end{eqnarray}
where ${\cal A}^p$ is the scattering amplitude of the $\gamma + p \to \Lambda+K^+$ reaction; ${\cal A}^n$ is the scattering amplitude of the $\gamma + n \to \Lambda+K^0$ 
reaction~\cite{MAID}.

Finally, the nuclear suppression factors $S_D(t,p_n)$ and $S_D(t)$ are defined
by the following expressions
\begin{eqnarray}
S_D(t,p_n)&=& \left(\frac{\sqrt{(2 \pi)^3 2 m_N}}{16 \pi}\right)^2 
\nonumber\\
&\times& 
\int dk \,
\frac{k}{E_k}\Theta(E_{\Theta}-E_k)\frac{\theta(E_{\gamma},t,M^2_{NK};k)}{p_{\Theta}} \frac{p_n}{E_n}
 \frac{\theta(E_{\gamma},t,M^2_{NK};p_n)}{p_{\Theta}}\rho_D(k,p_n) \,, \nonumber \\
S_D(t)&=&\left(\frac{\sqrt{(2 \pi)^3 2 m_N}}{16 \pi}\right)^2\nonumber\\
&\times& \int dk \,dk^{\prime} 
\frac{k}{E_k}\Theta(E_{\Theta}-E_k)\frac{\theta(E_{\gamma},t,M^2_{NK};k)}{p_{\Theta}}
\nonumber\\
 &\times& \frac{k^{\prime}}{E_{k^{\prime}}}\Theta(E_{\Theta}-E_{k^{\prime}}) \frac{\theta(E_{\gamma},t,M^2_{NK};k^{\prime})}{p_{\Theta}} 
\rho_D(k,k^{\prime}) \,, 
\end{eqnarray}
where $\rho_D$ is the unpolarized deuteron density matrix, which is expressed 
in terms of the $S$ and $D$-waves of the non-relativistic deuteron
wave function
\begin{eqnarray}
\rho_D(k,k^{\prime}) & = & u(k) u(k^{\prime})+w(k) w(k^{\prime})\left(\frac{3}{2}
\frac{(\vec{k} \cdot \vec{k}^{\prime})^2}{|\vec{k}|^2 \,|\vec{k}^{\prime}|^2 }-\frac{1}{2}\right) \,, \nonumber\\
&=& u(k) u(k^{\prime})+w(k) w(k^{\prime})\,.
\end{eqnarray}
The last line is a consequence of the fact that the delta-functions in the
loop integral over the spectator momenta in Eq.~(\ref{eq:a_signal2})
make the vectors $\vec{k}$
and $\vec{k}^{\prime}$ collinear.

Examples of the $d \sigma_{\Lambda}^{p+n}/dt$ and $d \sigma_{\Lambda}^{p}/dt$
cross sections and the $S(t)$ nuclear suppression factor can be found in our 
original publication~\cite{Guzey:2004jq} and shall not be repeated here.

In our original work~\cite{Guzey:2004jq}, we mostly concentrated on 
the triple differential cross section 
$d \sigma/(dt d p_n d M^2_{NK})$ at large $p_n$ ($p_n \geq 0.3$ GeV),
 which is dominated by the signal contribution. In the present work, 
in order to better compare our predictions to the results of the
CLAS analysis~\cite{Niccolai:2006td}, we concentrate on the $t$-integrated cross section $d \sigma/(d p_n d M^2_{NK})$ as a function of $M_{NK}^2$ and $p_n$
for $0.1 \leq p_n \leq 0.3$ MeV.
We shall also consider the $p_n$-integrated cross section.

Before performing any numerical calculations, it is important to qualitatively
understand the behavior of $d \sigma/(d p_n d M^2_{NK})$.
For the values of the invariant mass of the $nK^+$ system, $M_{nK}$,
which is away from the expected mass of the $\Theta^+$, 
$M_{\Theta}=1540$ MeV, by more than approximately 5 MeV, the $d \sigma^{{\rm I}}/(d p_n d M^2_{NK})$ and $d \sigma^{{\rm S}}/(d p_n d M^2_{NK})$ cross sections
are negligibly small. The resulting $d \sigma/(d p_n d M^2_{NK})$ is a
smooth function of $M_{nK}$, see Fig.~4 of Ref.~\cite{Niccolai:2006td}.
When $M_{nK}$ is close to the expected $\Theta^+$ peak,
$|M_{nK}-M_{\Theta}| \leq 1-5$ MeV, one expects a  very narrow peak, whose
magnitude depends on $p_n$ and does not depend on $\Gamma_{\Theta}$.

Next we discuss 
the $p_n$-dependence of $d \sigma/(d p_n d M^2_{NK})$ in the vicinity of the
$\Theta^+$ peak. The background, interference and signal cross section
have very distinct $p_n$-dependencies, which are determined by the 
corresponding nuclear suppression factors.
The background cross section is the largest of three in magnitude at small 
$p_n$, but it rapidly decreases with increasing $p_n$ due to the 
$|\psi_D(p_n)|^2$ factor. The interference cross section is smaller in 
magnitude than the background cross section because of the small 
value of
$\Gamma_{\Theta}$, but its decrease with increasing $p_n$ is not so steep,
which is roughly determined by the $\psi_D(p_n)$ factor in $S(t,p_n)$.
The signal cross section is the smallest of three in magnutude, but this
smallness is compenstated by the slow $p_n$-dependence.

At small values of the final neutron momenta,
$p_n \leq 0.05$ GeV, the background dominates the cross section and
one does not expect any resonance structures over the smooth
background.
As $p_n$ increases, $0.1 \leq p_n \leq 0.2-0.3$ GeV, the 
interference cross section becomes sizable, which results in a very narrow
negative peak superimposed on the smooth background.
As $p_n$ is increased further, $p_n \geq 0.3-0.4$ GeV, the signal 
cross section exceeds the background and the interference cross sections, which results
in a prominent peak over the background.

\begin{figure}[th]
\begin{center}
\epsfig{file=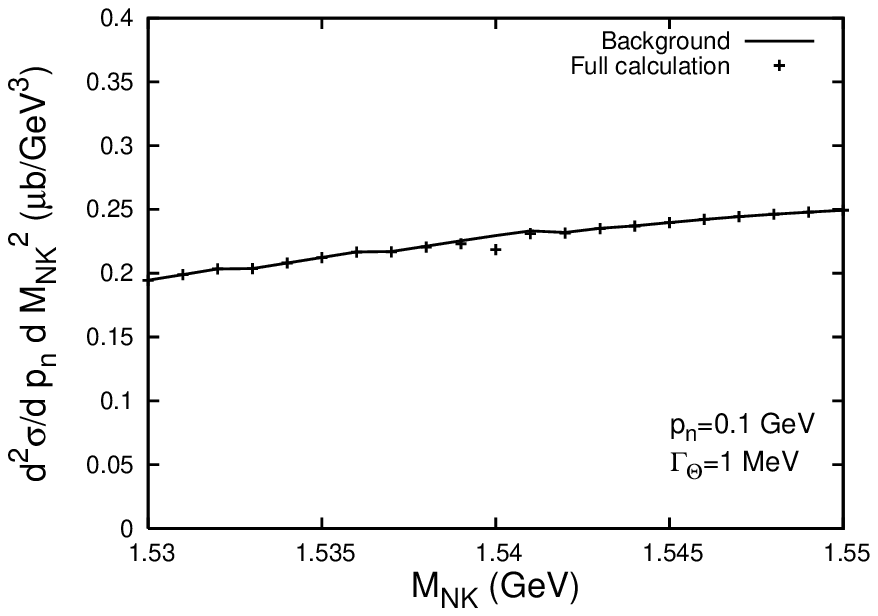,width=7cm,height=5.5cm}
\epsfig{file=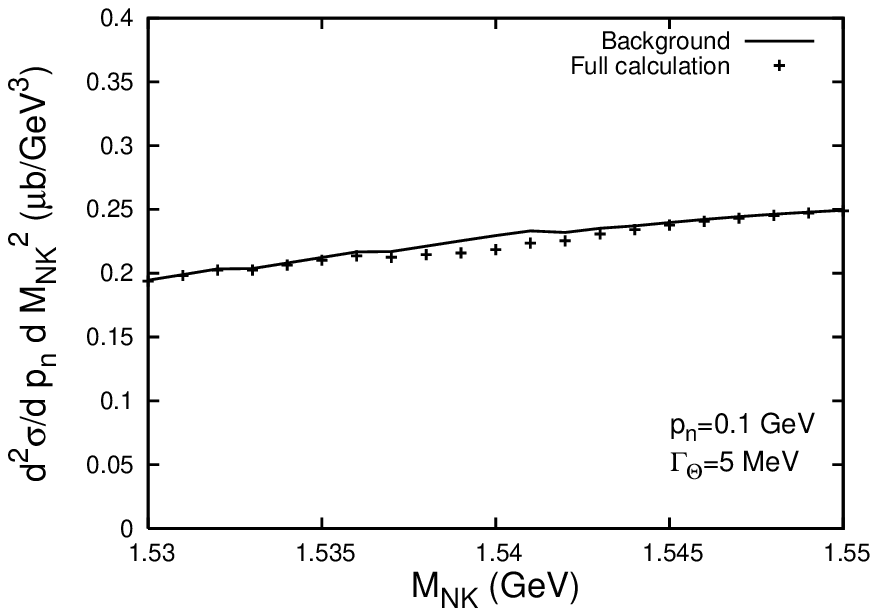,width=7cm,height=5.5cm}
\epsfig{file=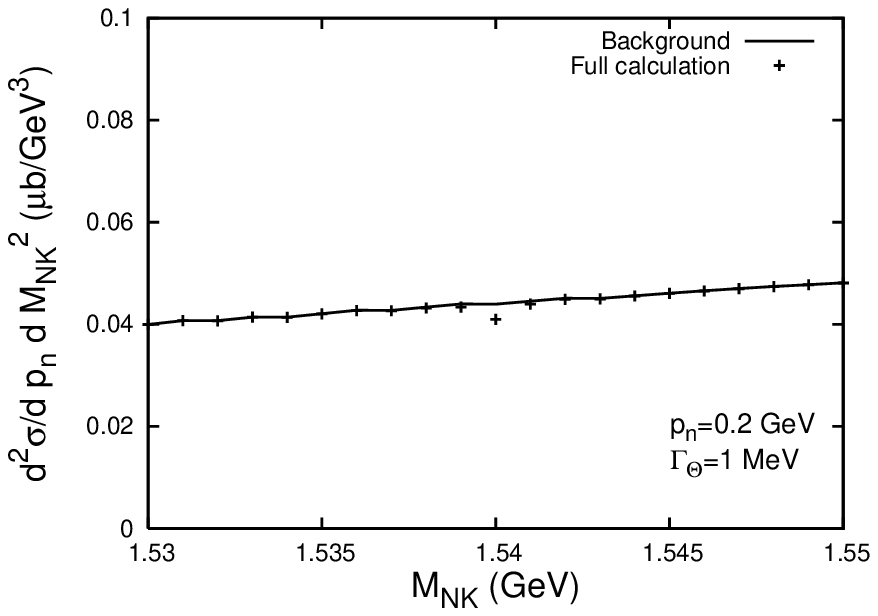,width=7cm,height=5.5cm}
\epsfig{file=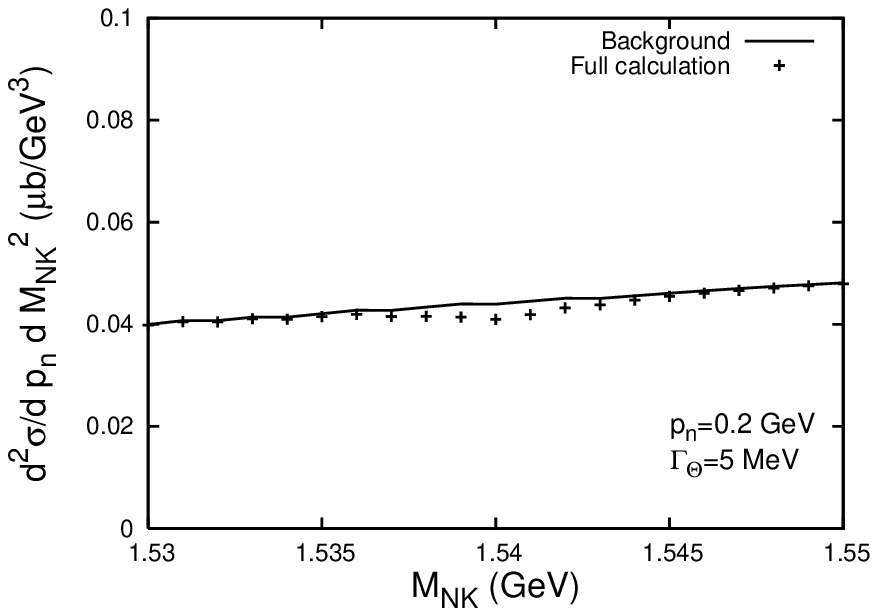,width=7cm,height=5.5cm}
\epsfig{file=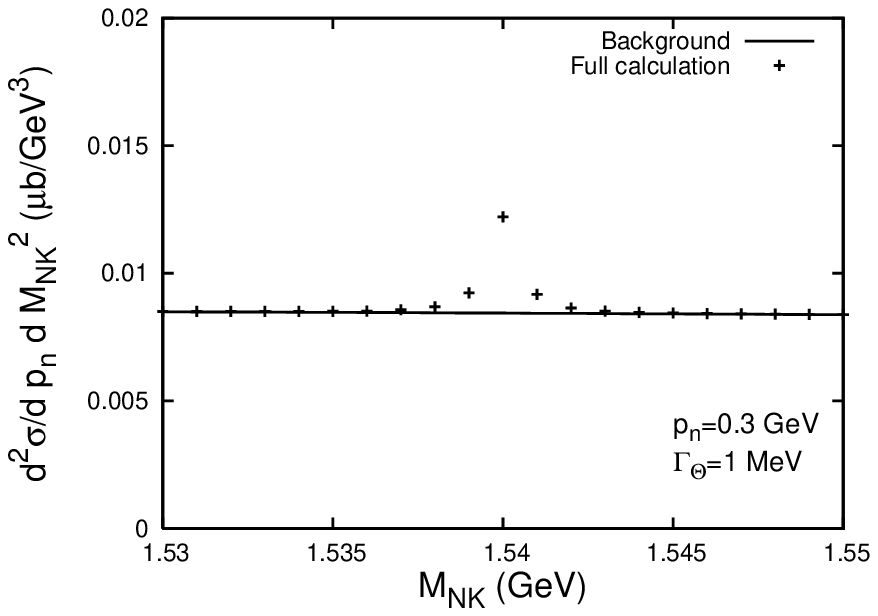,width=7cm,height=5.5cm}
\epsfig{file=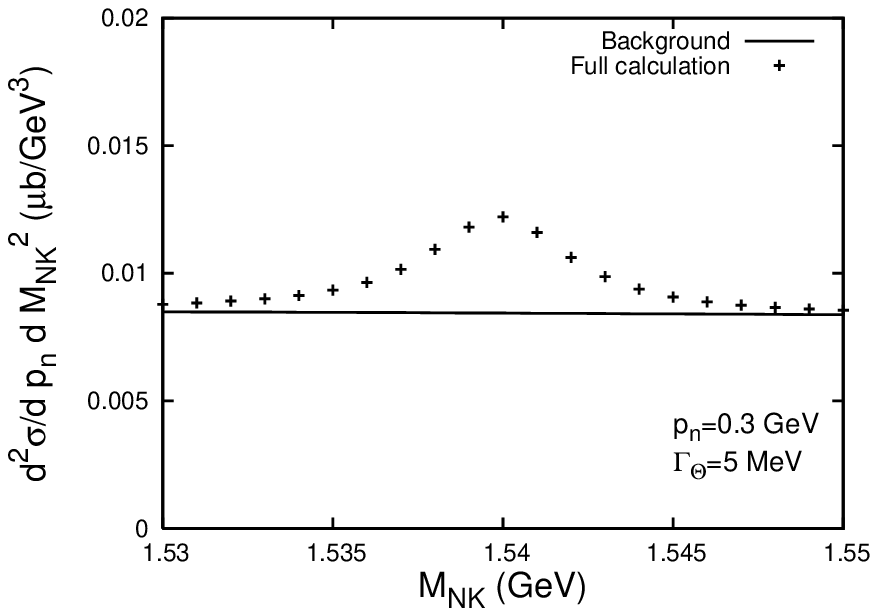,width=7cm,height=5.5cm}
\caption{The $d^2 \sigma/(d p_n d M^2_{NK})$ cross section of the
$\gamma+D \to \Lambda+n+K^+$ reaction at $E_{\gamma}=1.2$ GeV.
The solid curves is the background contribution; the crosses is the full
result.}
\label{fig:result12}
\end{center}
\end{figure}

This behavior is demonstated in Fig.~\ref{fig:result12}, where we plot
the double differential cross section $d \sigma/(d p_n d M^2_{NK})$ as
a function of $M^2_{NK}$ at different values of $p_n$. The energy of
the photon beam is $E_{\gamma}=1.2$ GeV, where the elementary $\gamma+N \to
\Lambda+K$ cross section passes through its maximum.
In order to examine the dependence on the total width of the 
$\Theta^+$, we performed calculation with $\Gamma_{\Theta}=1$ MeV
(left column)
and $\Gamma_{\Theta}=5$ MeV (right column).
The solid curves represent the background cross section $d \sigma^{{\rm BG}}/(d p_n d M^2_{NK})$;
the crosses represent the full cross section $d \sigma/(d p_n d M^2_{NK})$.

Figure~\ref{fig:result12} clearly depicts the interplay between the
background, interference and signal cross sections described above.
Note that as follows from Eq.~(\ref{eq:main}), the hight of the $\Theta^+$
signal does not depend on $\Gamma_{\Theta}$. 

\begin{figure}[th]
\begin{center}
\epsfig{file=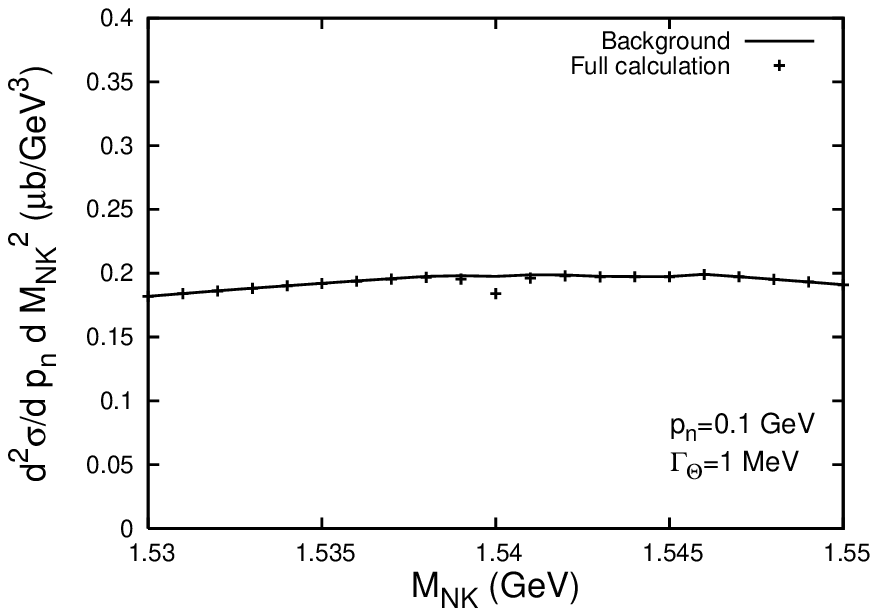,width=7cm,height=5.5cm}
\epsfig{file=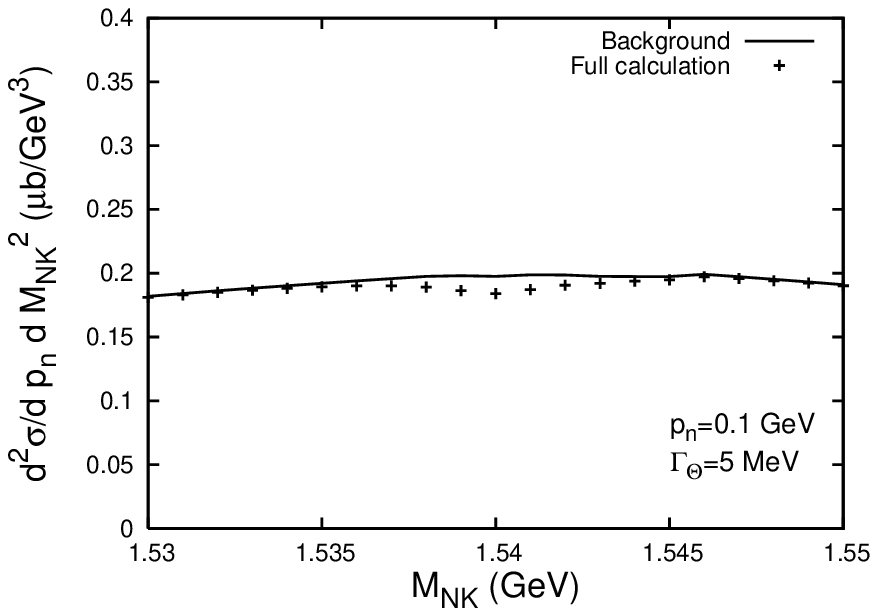,width=7cm,height=5.5cm}
\epsfig{file=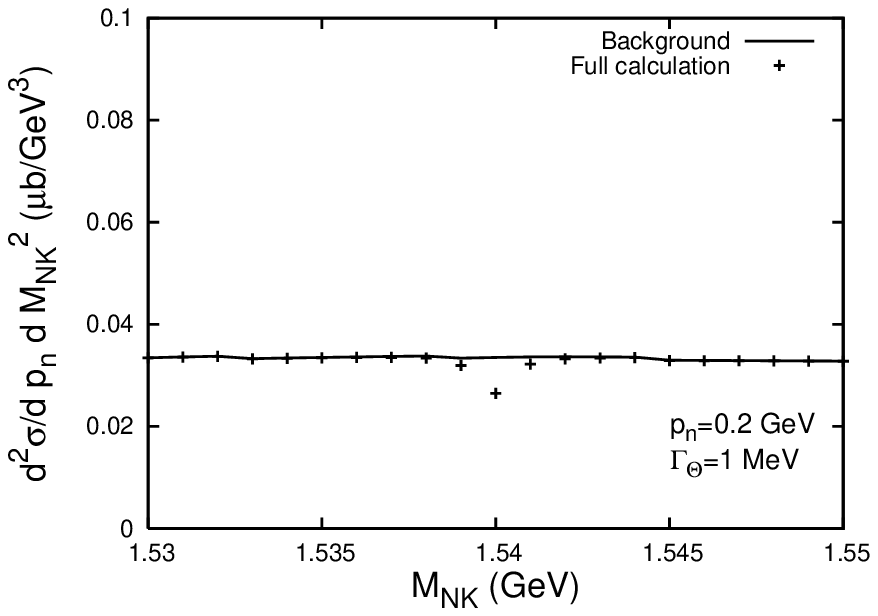,width=7cm,height=5.5cm}
\epsfig{file=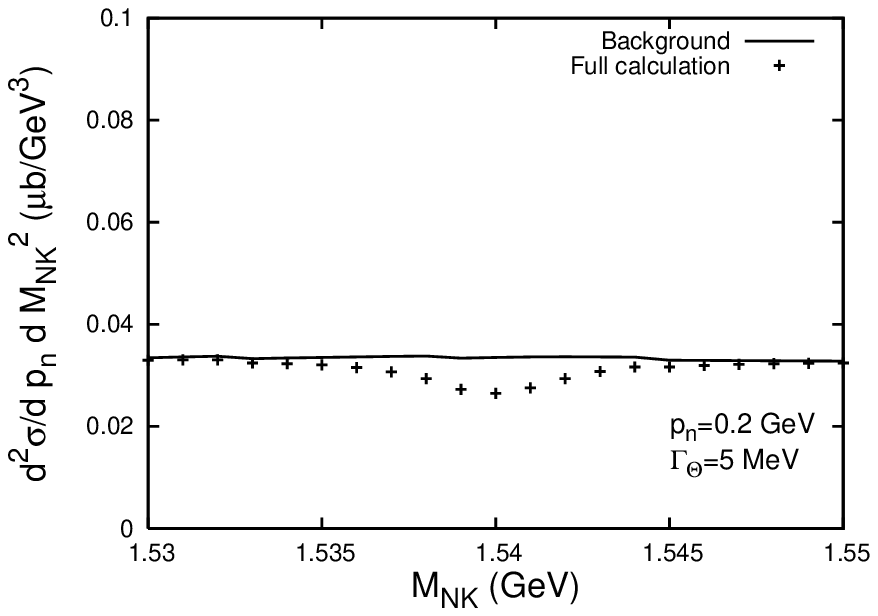,width=7cm,height=5.5cm}
\epsfig{file=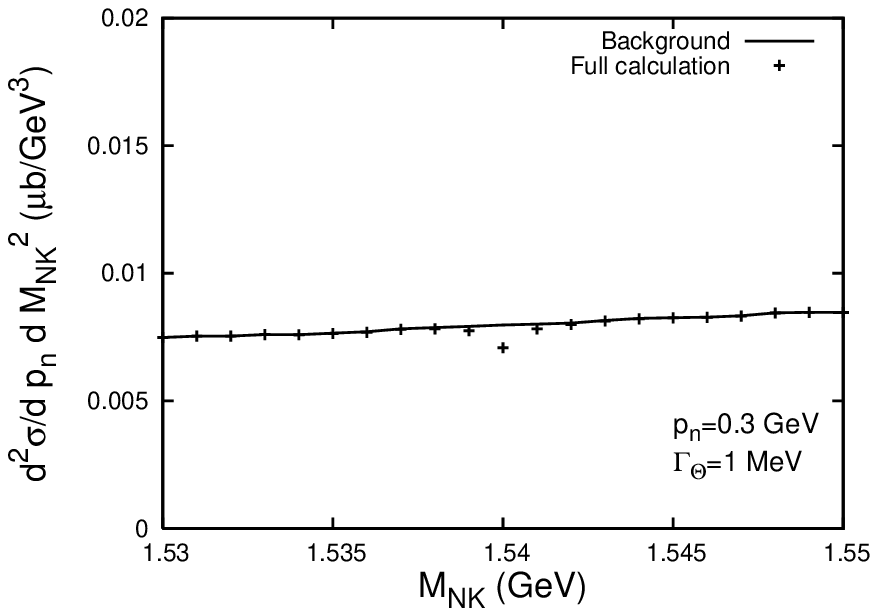,width=7cm,height=5.5cm}
\epsfig{file=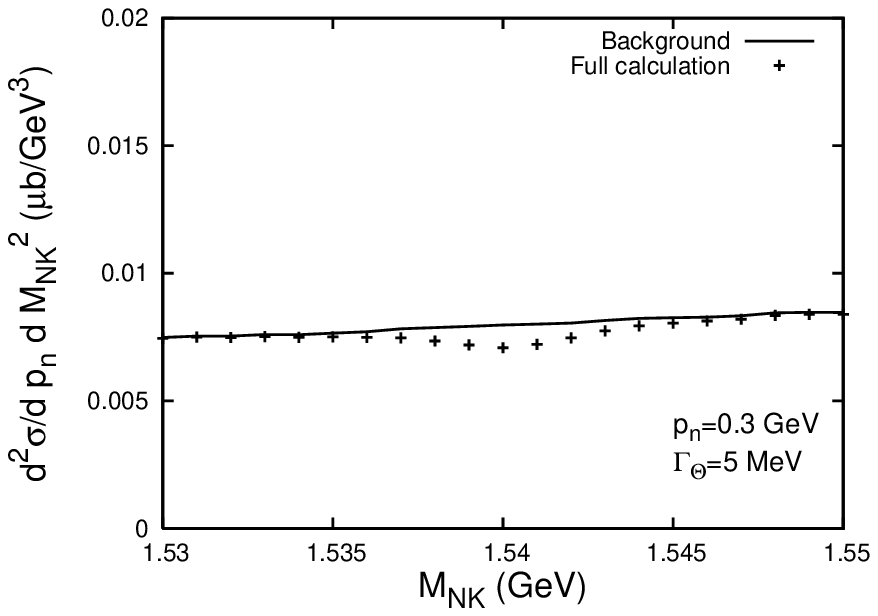,width=7cm,height=5.5cm}
\caption{The $d^2 \sigma/(d p_n d M^2_{NK})$ cross section of the
$\gamma+D \to \Lambda+n+K^+$ reaction at $E_{\gamma}=1.6$ GeV. 
The solid curves is the background contribution; the crosses is the full
result.}
\label{fig:result16}
\end{center}
\end{figure}

The combinations of the elementary $\gamma+N \to \Lambda+K$ cross 
sections~(\ref{eq:lambda}) depend on the photon energy.
Figure~\ref{fig:result16} presents our predictions for the $d \sigma/(d p_n d M^2_{NK})$ cross section calculated with $E_{\gamma}=1.6$ GeV.
Our results presented in Figs.~\ref{fig:result12} and \ref{fig:result16}
are rather similar for $p_n=0.1$ GeV and $p_n=0.2$. However, 
at $E_{\gamma}=1.6$ GeV and $p_n=0.3$ GeV, as a consequence
of the cancellation between the ${\cal A}^p$ and ${\cal A}^n$ amplitudes,
the interference cross section is 
still larger than the signal cross section. For larger $p_n$, $p_n \geq 0.4$ GeV,
the signal cross section will win over the interference cross section and 
one will observe a distinct peak, similarly to the lower panels of Fig.\ref{fig:result12}.

The size of the $\Theta^+$ signal
(the deviation of the crosses from the smooth background)
depends on the photon energy $E_{\gamma}$. At $E_{\gamma}=1.2$ GeV, the 
deviation is 5-7\% for $p_n=0.1-0.2$ GeV and rapidly becomes as large as
45\% at $p_n=0.3$ GeV. At  $E_{\gamma}=1.6$ GeV, the deviation is
7\% at $p_n=0.1$ GeV, 27\% -- at $p_n=0.2$ GeV,  and it is 12\% at  
 $p_n=0.3$ GeV.

While the numbers for the $\Theta^+$ signal look impressive, the following
analysis will demonstrate that if the $\gamma+D \to \Lambda+n+K^{+}$ differential 
cross section is integrated over $t$, $p_n$ and $E_{\gamma}$ as was done in 
the CLAS analysis~\cite{Niccolai:2006td}, the $\Theta^+$ signal disappears,
mostly as a result of the cancellation between the interference and the signal
contributions.

In detail, in order to compare our predictions to the CLAS result~\cite{Niccolai:2006td}, we integrate the  $\gamma+D \to \Lambda+n+K^{+}$ differential cross section of Eqs.~(\ref{eq:kajantie4}) and (\ref{eq:main})
over $t$ and $p_n$. In order to simulate the lack
of the forward acceptance of the CLAS detector, the minimal value of $t$ in the
integration over $t$ is set by the cut on the $\Lambda$ scattering angle,
$\cos \theta_{\Lambda} \leq 0.9$.
The integration over the final neutron momentum $p_n$ is performed in the 
intervals $0 \leq p_n \leq 0.6$ GeV and $0.2 \leq p_n \leq 0.6$ GeV. The upper
limit of integration is chosen somewhat arbitrary, with the aim to reduce the 
sensitivity of our results to the poorly known high-momentum tail of the 
deuteron wave function and to possible relativistic corrections.
Since we are mostly interested in the hight of the $\Theta^+$ signal and not
in its shape, this
calculation is performed with $\Gamma_{\Theta}=1$ MeV.

\begin{figure}[th]
\begin{center}
\epsfig{file=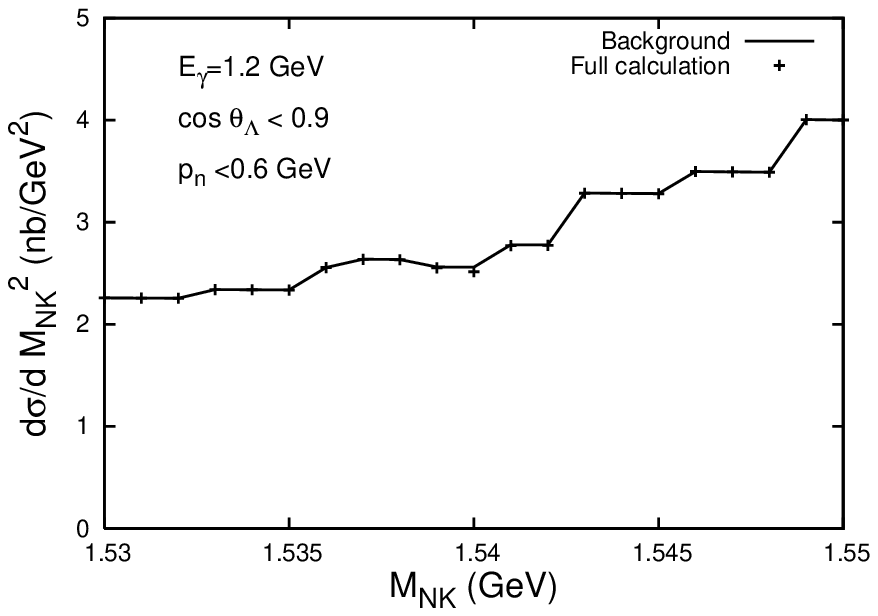,width=8cm,height=8cm}
\epsfig{file=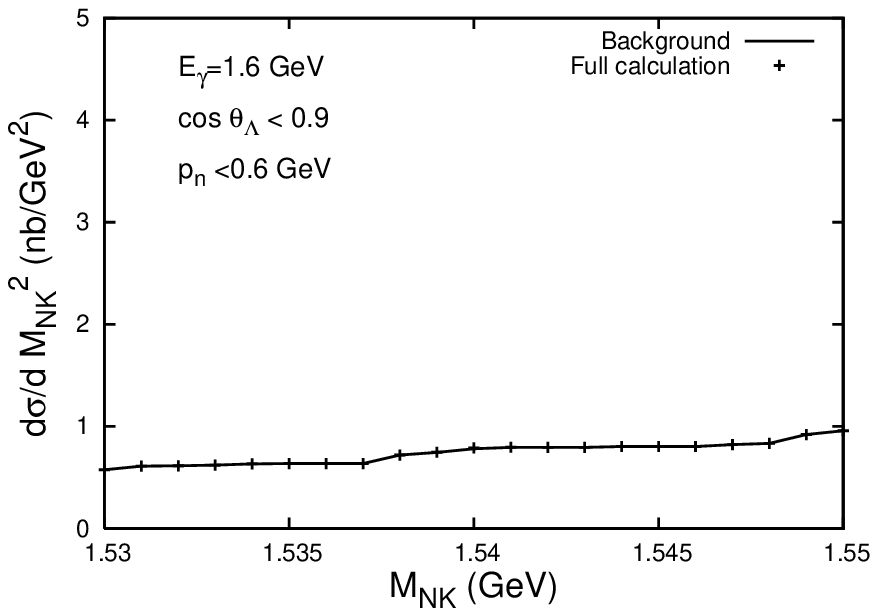,width=8cm,height=8cm}
\epsfig{file=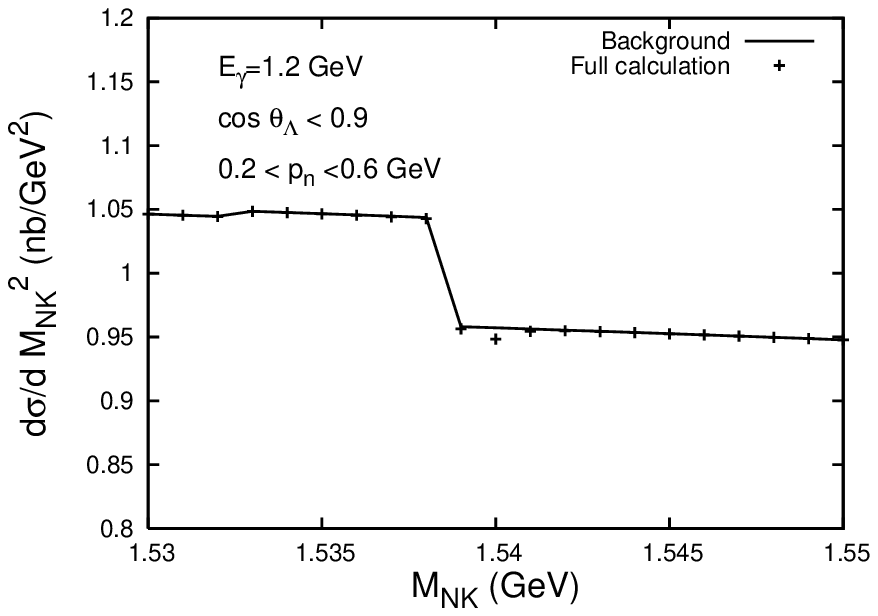,width=8cm,height=8cm}
\epsfig{file=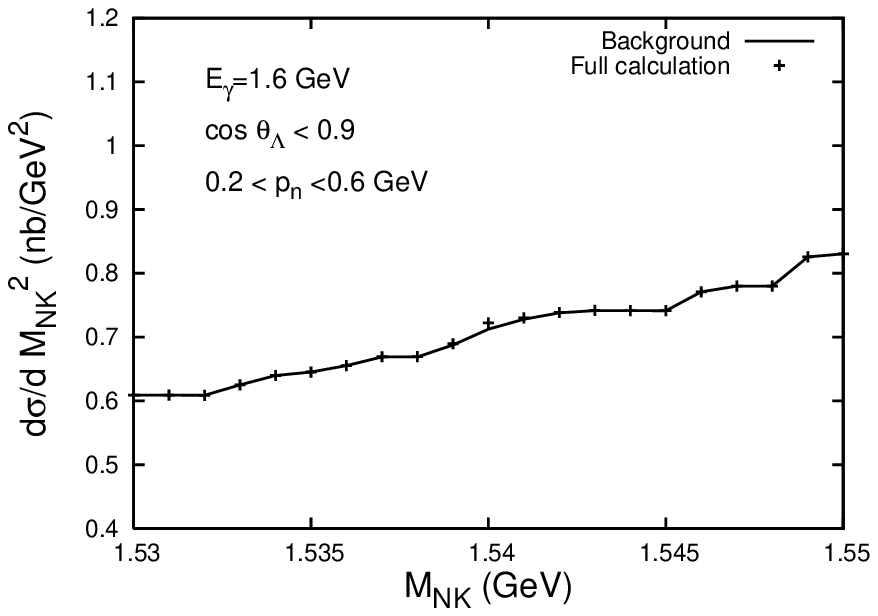,width=8cm,height=8cm}
\caption{The $d \sigma/d M^2_{NK}$ cross section of the
$\gamma+D \to \Lambda+n+K^+$ reaction at $E_{\gamma}=1.2$ GeV and $E_{\gamma}=1.6$ GeV.
The solid curves is the background contribution; the crosses is the full
result. The calculation uses $\Gamma_{\Theta}=1$ MeV.}
\label{fig:result_ps_integrated}
\end{center}
\end{figure}

Our results for two photon energies, $E_{\gamma}=1.2$ GeV and $E_{\gamma}=1.6$ GeV,
are presented in Fig.~\ref{fig:result_ps_integrated}. The upper panels correspond
to the integration over the $0 \leq p_n \leq 0.6$ GeV interval;
the lower panels correspond to the $0.2 \leq p_n \leq 0.6$ GeV interval.
The solid curves represent the background contribution; the crosses give the 
results of the full calculation. 
One should not the change of the scale from $\mu$b in Fig.~\ref{fig:result12} and \ref{fig:result16} to nb in Fig.~\ref{fig:result_ps_integrated}, which
is responsible for the non-smooth curves in the latter figure.

Two features of Fig.~\ref{fig:result_ps_integrated} deserve a discussion.
First, the results of the calculations with $E_{\gamma}=1.2$ GeV and $E_{\gamma}=1.6$ GeV
are fairly different. This is a consequence of the $E_{\gamma}$-dependence
of the elementary $\gamma+N \to \Lambda +K$ amplitudes. Second, 
the cut on the minimal value of the
final neutron momentum~\cite{Guzey:2004jq} is intended in order
to enhance the $\Theta^+$ signal. Indeed, a comparison 
between the upper left  and lower left panels shows that a tiny negative peak
at $M_{nK}=1.540$ GeV appears. The deviation of the peak from the background is
only 2\%. Similarly for $E_{\gamma}=1.6$ GeV, a comparison 
between the upper right  and lower right panels shows that a tiny 1\% positive peak
appears. Clearly, it is impossible to experimentally observe such tiny
deviations from the background. 

In summary, while the cut on the minimal value of 
the final neutron momentum indeed enhances the 
$\Theta^+$ signal, the cancellation between the interference and signal 
contributions in the process of integration over $p_n$ completely washes out
the desired $\Theta^+$ signal. 
It appears that the $\Theta^+$ signal from the $\gamma+D \to \Lambda+n+K^+$ cross section
can be successfully extracted only if tighter cuts on $p_n$ and
$t$ are imposed, see examples in Ref.~\cite{Guzey:2004jq} and
Figs.~\ref{fig:result12} and \ref{fig:result16}
of this work.

\section{Conclusions and discussion}

We extended our original analysis of the $\Theta^+$
production in the $\gamma+D \to \Lambda +n +K^+$ reaction~\cite{Guzey:2004jq}
in order to provide a better comparison to the recent 
CLAS measurement~\cite{Niccolai:2006td}.
We studied the dependence of the $\gamma+D \to \Lambda +n +K^+$ differential
cross section on the $n K^+$ invariant mass and on the momentum of the
final neutron, $p_n$. We demonstrated the important role 
of the interference between the 
signal and background contributions to the $\gamma+D \to \Lambda +n +K^+$
amplitude 
and examined the $p_n$-dependence of the resulting signal, interference and
background cross sections. This allowed us to identify kinematic conditions 
providing the enhanced sensitivity to the $\Theta^+$ signal.

We attempted to perform a realistic comparison to the results of the CLAS 
measurement~\cite{Niccolai:2006td} by integrating the $\gamma+D \to \Lambda +n +K^+$
differential cross section over $t$ and $p_n$. 
The particular choice of cuts implemented in the integration over $t$ and
$p_n$ consists the main source of theoretical uncertainty of the results presented in
Fig.~\ref{fig:result_ps_integrated}.
We showed that as a result of the
cancellation between the interference and signal contributions, the
$\Theta^+$ signal almost completely washes out after the integration over $p_n$, even
when the $p_n > 0.2$ GeV cut is imposed.
This is consistent with the CLAS conclusion that no statistically 
significant structures were observed~\cite{Niccolai:2006td}.

Therefore, there is no disagreement between the theory and the experiment
and
the CLAS result does not refute the existence of the $\Theta^+$.

It appears that the $\Theta^+$ signal from the $\gamma+D \to \Lambda+n+K^+$ cross section
can be successfully extracted only if tighter cuts on $p_n$ and
$t$ are imposed, see Figs.~\ref{fig:result12} and \ref{fig:result16}.
In this respect, it seems interesting to analyse further the small
devitation from the smooth background near $M_{nK} \approx 1.53$ GeV seen in the 
lower panel of Fig.~4 of the CLAS publication~\cite{Niccolai:2006td}.

\acknowledgments

We thank M.~Polyakov for noticing the incorrect sign of the interference 
cross section in our original publication~\cite{Guzey:2004jq}
 and for stimulating the present research.

\end{document}